\documentclass[english,a4paper,twocolumn]{revtex4}
\usepackage[T1]{fontenc}
\usepackage[latin1]{inputenc}
\usepackage{amsmath}
\usepackage{graphicx}
\usepackage{amssymb}
\usepackage{times}

\makeatletter

\makeatletter

\makeatletter

\makeatletter

\usepackage{bm}

\usepackage{amsfonts}

\makeatother

\makeatother

\makeatother

\usepackage{babel}
\makeatother
\begin{document}

\title{Quantum many-body simulations using Gaussian phase-space representations}

\author{P. D. Drummond, P. Deuar, J. F. Corney}

\affiliation{ARC Centre of Excellence for Quantum Atom Optics, University of Queensland,
Brisbane, Queensland, Australia.}

\begin{abstract}
Phase-space representations are of increasing importance as a viable
and successful means to study exponentially complex quantum many-body
systems from first principles. This paper traces the background of
these methods, starting from the early work of Wigner, Glauber and
Sudarshan. We focus on modern phase-space approaches using non-classical
phase-space representations. These lead to the Gaussian representation,
which unifies bosonic and fermionic phase-space. Examples treated
include quantum solitons in optical fibers, colliding Bose-Einstein
condensates, and strongly correlated fermions on lattices. 
\end{abstract}
\maketitle

\section{Introduction}

In this paper, we will trace how the concept of coherence and coherent
states has led an important advance: the quantum phase-space representation.
Through the development of phase-space representations, the idea of
coherence can help the understanding and simulation of the physics
of many-body systems, both in thermal equilibrium, and in time-dependent,
quantum dynamical calculations. This is of increasing importance beyond
quantum optics, as new experiments explore the quantum correlations
and dynamics of interacting particles.

We show that a more general approach to coherence leads to the Gaussian
phase-space method, which unifies the representation of both bosonic
and fermionic many-body systems. This powerful idea has many ramifications.
It encompasses all the known bosonic representations in a simple,
clear formalism, and extends these ideas to fermions as well. It is
also extremely useful in applications, as we will show using both
equilibrium and non-equilibrium examples.

A particular quantum state that illustrates this is the coherent state.
It was introduced originally by Schroedinger\cite{Schrodinger:1926}
for the harmonic oscillator, and later applied to the radiation field
through the seminal work of Sudarshan\cite{Sudarshan:1963} and Glauber\cite{Glauber:1963}.
These states are fully coherent in the sense that normally ordered
operator moments factorize to all orders.

The definition of a coherent state is extremely simple. If $\widehat{a}$
is a field-mode annihilation operator, then the coherent state is
defined as a normalized eigenstate of $\widehat{a}$,\begin{equation}
\widehat{a}\left|\alpha\right\rangle =\alpha\left|\alpha\right\rangle .\end{equation}
 These states form a complete mathematical basis, providing examples
of quantum states which are perfectly coherent to all orders. The
idea can be extended to other algebras, for example the SU(N) coherent
states, and were used to construct the P-representation - a representation
of the radiation field in terms of diagonal coherent state projection
operators. This \textit{quantum operator representation} has the
form (for a single mode) of:

\begin{equation}
\,\widehat{\rho}=\int P(\alpha)\left|\alpha\right\rangle \left\langle \alpha\right|d^{2}\alpha\,.\label{eq:P-function}\end{equation}
 This representation maps a quantum state into a distribution on a
classical phase space. Other representations like this exist, including
the Wigner\cite{Wigner1932a} representation and the Husimi\cite{Husimi:1940}
Q-function. The closely related operator associations of Lax\cite{Lax:1966},
Agarwal\cite{Agarwal:1970,Agarwal:1970a} and co-workers were used
to develop a quantum theory of the laser. While useful for the laser,
these all lack essential ingredients that would allow them to be useful
as a probability distributions in first-principles many-body dynamical
simulations. Most are simply non-positive, as in the case of the P-function
and Wigner function. Any representation that uses a classical-like
phase space has no corresponding exact stochastic equation when there
are inter-particle interactions.

We will explain how this problem is solved by extending the phase-space
dimension, giving rise to the positive P-representation\cite{Chaturvedi:1977a,Drummond1980a}.
A unifying principle is the use of non-orthogonal basis sets, which
leads to the idea of a stochastic gauge symmetry\cite{Deuar:2002},
and more general Gaussian phase-space methods\cite{Corney:2003,Corney2006b}.
These have many applications to interacting Bose and Fermi systems.
Both thermal equilibrium and first-principles quantum dynamical time-evolution
(either unitary or dissipative) can be treated. Recent bosonic examples
include quantitatively tested predictions on quantum soliton time-evolution\cite{Corney2006c},
as well as novel predictions for topical experiments including: colliding
Bose-Einstein condensates\cite{Deuar2006b}, tunnel-coupled condensates\cite{Poulsen2001b}, superchemistry
\cite{Hope2001a},  molecular
dissociation\cite{Poulsen2001a,Kheruntsyan2002,Savage2006,Kheruntsyan2005a}, micro-mechanical
resonators\cite{Olsen:2004}, triple EPR correlations\cite{Olsen:2006qy},
and non-equilibrium criticality in parametric downconversion\cite{Drummond:2005fk}.
We also give results for phase-space simulations of the fermionic
Hubbard model in thermal equilibrium\cite{Corney2004a}.

\section{Quantum Many-body systems}

Quantum many-body theory is the generic theory we currently use for
describing all non-astronomical physical systems from a microscopic
point of view. It is applicable to a wide range of problems.

\subsection{Ultra-cold atomic Gases}

As simple examples of interacting quantum systems, consider the ultra-cold
atomic Bose-Einstein condensates and degenerate atomic Fermi gases.
Ultra-cold atoms are an ideal quantum many-body system. In these experiments,
the interacting atoms are isolated from other matter, by virtue of
being optically or magnetically trapped in a high-vacuum environment
at low temperatures. Important advances in the last decade include:
Bose-Einstein condensation (BEC), atom lasers, superfluid Fermi atoms,
superchemistry (stimulated molecule formation), atomic diffraction,
interferometers, and temperatures below $1nK$.

Such well-controlled and simple physical systems present an opportunity
to quantitatively test quantum mechanics in new regimes, where macroscopic
and many-body effects play a dominant role.

\subsection{Many-body quantum dynamics}

Before one can make quantitative predictions, there is a significant
problem to overcome: quantum many-body problems are exponentially
complex.

To illustrate this, consider a Bose gas with $N$ atoms distributed
among $M$ modes. Each mode can have one or all atoms. The number
$N_{s}$ of quantum states available is:\begin{equation}
N_{s}=\frac{\left(N+M-1\right)!}{N!\left(M-1\right)!}\,.\end{equation}
 A typical BEC may have $N\simeq M\simeq500,000$, giving the astronomical
number of:\begin{equation}
N_{s}=2^{2N}=10^{300,000}\,.\end{equation}
 Hilbert space dimension can also be classified by the number of equivalent
quantum bits (qubits), which is $\log_{2}N_{s}=2N=1,000,000$, in
this example.

There are a number of possible solutions to dynamical problems. Here
we focus on methods which are exact, in the sense that errors can
be estimated and reduced where necessary. As an example, while Density
Matrix Renormalisation Group (DMRG) methods\cite{Schollwock2005}
can be useful for one-dimensional calculations, including dynamics,
the Hilbert-space truncation is not always a well-controlled approximation.
Similar difficulties occur in the density functional approach\cite{Kohn1999}.
Uncontrolled approximations cannot be used as a basis for \emph{testing}
quantum mechanics. Any discrepancies observed may simply be caused
by calculational errors, rather than fundamental issues.

Candidates for \emph{exact} solutions are as follows:

\begin{description}
\item [{{Path~integrals~and~Monte-Carlo}}] - these are useful for
bosons at thermal equilibrium. For quantum dynamics and for fermions,
there are phase and sign problems, making these methods often impractical. 
\item [{{{Perturbation~theory}}}] - while applicable for certain problems,
this method generally doesn't converge in quantum field theory 
\item [{{{Numerical~diagonalization}}}] - the problem of an exponentially
large matrix size rules out such brute force methods, except for very
small particle numbers 
\item [{{{Exact~solutions}}}] - even if all the energy eigenstates
are known (which is unusual) evaluating the initial expansion coefficients
for quantum dynamics remains exponentially difficult, and therefore
impractical 
\item [{{{New~hardware}}}] - Feynman proposed quantum computers to
solve many-body problems - currently, these do \textbf{not} exist
beyond $2-4$ qubit capacity 
\item [{{{\emph{New~software}}}}] - \emph{Gaussian quantum phase-space
simulation methods can give practical techniques using existing computers,
simulating quantum systems equivalent to nearly a million qubits.}
\end{description}

\section{Quantum Phase-space methods}

The great power of phase-space methods is their ability to accurately
compute the quantum dynamics of fully macroscopic systems directly
from the Hamiltonian, without resorting to overarching approximations.
This confers several advantages over previous methods, despite the
introduction of randomness that limits precision:

\begin{description}
\item [{{{\textit{Firstly},}}}] all uncertainty in the results is confined
to random statistical fluctuations, with no systematic bias. Importantly,
the magnitude of this uncertainty can be reliably estimated from the
distribution of sub-ensemble means by using the central limit theorem 
\item [{{{\textit{Secondly},}}}] these methods lead to relatively simple
equations that can be easily adapted to trap potentials and local
losses, whose magnitude and shape can be chosen arbitrarily. This
is in stark contrast to approximate methods, which can become much
more complicated or even inapplicable under such conditions. 
\end{description}
The Gaussian quantum phase-space representation described here encompass
all the earlier known phase-space methods. Therefore, we start by
reviewing these earlier approaches.

\subsection{Classical and Quantum phase space }

Wigner\cite{Wigner1932a} originated the idea of a classical-like
phase-space or quasi-probability description. For $M$ modes, these
methods scale linearly with mode number, having just $M$ complex
dimensions. Variations on this theme include the Husimi Q-function\cite{Husimi:1940},
and the Glauber-Sudarshan\cite{Glauber:1963,Sudarshan:1963} P-representation.
For many quantum states, they result in a positive-valued distribution.
For the Q-function, this is always true. Despite this, one finds that
there is no corresponding stochastic equation for cubic or quartic
Hamiltonians. Thus, there is no method for efficiently time-evolving
a sampled distribution of an interacting system, except through an
approximate truncation of the equations of motion.

The solution to this problem is to use an enlarged phase space, which
includes off-diagonal terms in a coherent-state expansion. Intuitively,
this allows for quantum superpositions between more than one classical
configuration. The simplest possibility is the positive-P (+P) distribution\cite{Drummond1980a},
which has $2M$ coordinates. It results in a distribution function
which is always positive, and given certain conditions, obeys a stochastic
equation. It has the definition that:

\begin{equation}
\,\widehat{\rho}=\int P(\alpha,\beta)\frac{\left|\alpha\right\rangle \left\langle \beta^{*}\right|}{\left\langle \beta^{*}\right|\left|\alpha\right\rangle }d^{2}\alpha d^{2}\beta\,.\label{eq:+P-function}\end{equation}

\subsection{Quantum phase-space representations}

Guided by the formalism of Equations (\ref{eq:P-function}) and (\ref{eq:+P-function}),
one can define a general quantum phase-space representation by expanding
the density matrix $\widehat{\rho}$ using a complete basis of operators
\textbf{$\,\widehat{\Lambda}(\overrightarrow{\lambda})$:}

\begin{equation}
\,\widehat{\rho}=\int P(\overrightarrow{\lambda})\widehat{\Lambda}(\overrightarrow{\lambda})d\overrightarrow{\lambda}\,.\end{equation}
 Provided $P(\overrightarrow{\lambda})$ remains positive and sufficiently
bounded, quantum dynamics can be transformed into trajectories in
$\overrightarrow{\lambda}$. Different basis choices for $\,\widehat{\Lambda}(\overrightarrow{\lambda})$
then result in different representations. For example, the P-representation
has a single complex dimension (for $M=1$), so $\lambda_{1}=\alpha$,
and:\begin{equation}
\,\widehat{\Lambda}(\alpha)=\left|\alpha\right\rangle \left\langle \alpha\right|\,.\end{equation}

\begin{figure}
\includegraphics[width=6cm]{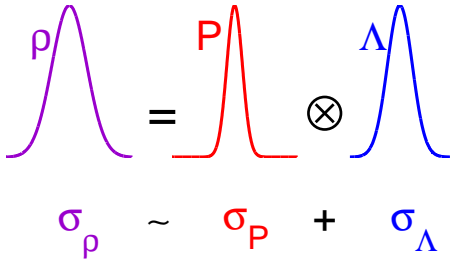}

\caption{\label{fig1} The full variance $\sigma_{\rho}$ is composed of a
distribution variance $\sigma_{P}$, and a basis variance $\sigma_{\Lambda}$.}
\end{figure}

As shown in Figure (\ref{fig1}), there are trade-offs in the choice
of basis, since the quantum variance is partly due to the distribution,
and partly due to the basis. By minimizing the the distribution variance,
one can reduce the sampling error of the representation. This typically
involves an over-complete, non-orthogonal basis in which each member
of the basis is closely matched to a physical state that occurs in
the simulation.

\subsection{Fermionic phase space}

Coherent states for fermions\cite{Martin1959a,Ohnuki1978a} can be
defined by means of anti-commuting Grassmann numbers, and have been
used, for example, in path-integral formulations for fermions\cite{Negele1989a}.
Like their bosonic counterparts, fermionic coherent states provide
an overcomplete basis set, and as Cahill and Glauber showed, can be
used to defined phase-space representations for fermions\cite{Cahill1999a}.
Unlike their bosonic counterparts, the fermionic coherent states have
no direct physical meaning. Moreover, while they are useful for formal
calculations, they have limited applicability as a basis for practical,
numerical calculations, because the of the complexity that arises
from the anticommuting properties of the algebra\cite{Plimak2001a}.
The coherent-state P-representation is then a function of Grassmann
numbers, not a probability.

But this anticommuting complexity is related to the unphysical states
contained in the coherent basis. Fermionic coherent states require
Grassmann numbers because of the way they include coherences between
states with an odd number difference. Consider a coherent superposition
of zero and one-particle states:$\left|\psi\right\rangle =\alpha\left|0\right\rangle +\beta\left|1\right\rangle $,
which gives a nonzero value for the coherence $\left\langle a\right\rangle =\alpha^{*}\beta$.
Because the $\left|1\right\rangle $ state involves an anticommuting
operator, one of the amplitudes must also be anticommuting, for consistency.
This also means that the coherent amplitude is anticommuting.

However, from superselection rules, we know that fermions can only
be created in pairs, and thus such superpositions are excluded. Thus
one can avoid this anticommuting problem by considering an operator
basis which only includes coherences that are allowed by the superselection
rule.

\subsection{General $M$-mode Gaussian operator }

The most general phase-space representation, for both fermions \emph{and}
bosons, is obtained with Gaussian operators. These provide an (over)complete
basis for fermions even when the coherences, and thus Grassmann components
are excluded\cite{Corney2006a}. These also generalize the concept
of coherence: physical states with Gaussian density operators have
operator products that factorize in a similar, but more general way
than coherent states.

To define these, we introduce $\widehat{\bm a}$ as a column vector
of $M$ bosonic/fermionic annihilation operators (indicated as the
upper or lower sign respectively), and $\widehat{\bm a}^{\dagger}$
the corresponding row vector of creation operators,. Their commutation
relations are:\begin{equation}
\left[\widehat{a}_{k},\widehat{a}_{j}^{\dagger}\right]_{\mp}=\delta_{kj}\,\,.\label{eq:a-com}\end{equation}

A Gaussian operator is defined as a normally ordered exponential of
a quadratic form in annihilation and creation operators. Introducing
extended $2M$-vectors of operators: $\underline{\widehat{a}}=(\widehat{\bm a},(\widehat{\bm a}^{\dagger})^{T})$,
with adjoint defined as $\underline{\widehat{a}}^{\dagger}=(\widehat{\bm a}^{\dagger},\widehat{\bm a}^{T})$,
the operator fluctuation is then: $\delta\underline{\widehat{a}}=\underline{\widehat{a}}-\underline{\alpha}$
, where $\underline{\alpha}=(\bm\alpha,\bm\beta^{*})$ is a $2M$-vector
c-number. A Gaussian operator can therefore be written as:

\begin{eqnarray}
\widehat{\Lambda}_{\pm} & = & \Omega\ \left|\underline{\underline{\sigma}}\right|^{\mp1/2}:\exp\left[\delta\underline{\widehat{a}}^{\dagger}\left(\underline{\underline{I}}_{f}-\frac{1}{2}\underline{\underline{\sigma}}^{-1}\right)\delta\underline{\widehat{a}}\right]:\,.\label{eq:Gaussbasis}\end{eqnarray}
 In the fermionic case the square root of the determinant (for normalization
purposes) is to be interpreted as the Pfaffian of the matrix, in an
explicitly antisymmetric form. The additional factor $\underline{\underline{I}}$
in the exponent only appears in the fermionic case: \begin{eqnarray}
\underline{\underline{I}} & \equiv & \left[\begin{array}{cc}
\pm\mathbf{I} & \mathbf{0}\\
\mathbf{0} & \mathbf{I}\end{array}\right]\,.\label{eq:constantmatrix}\end{eqnarray}

\subsection{Operator mappings}

The covariance $\underline{\underline{\sigma}}$ is best thought of
as a kind of dynamical Green's function. It can be expanded as: \begin{eqnarray}
\underline{\underline{\sigma}} & = & \left[\begin{array}{cc}
\pm\widetilde{\mathbf{n}}^{T} & \mathbf{m}\\
\mathbf{m}^{+} & \widetilde{\mathbf{n}}\end{array}\right]\,.\label{eq:sigma}\end{eqnarray}
 Here $\mathbf{n}$ is a complex matrix whose average is the normal
Green's function for particles, while $\widetilde{\mathbf{n}}\equiv\mathbf{1}\pm\mathbf{n}$.
In many-body terminology, $\mathbf{m}$ and $\mathbf{m}^{+}$ correspond
to anomalous Green's functions. The representation phase space is
therefore $\,\overrightarrow{\lambda}=(\Omega,\bm\alpha,\bm\beta,\mathbf{n},\mathbf{m},\mathbf{m}^{+})$
for bosons; in the case of fermions, one must set $\bm\alpha=\bm\beta=0$.

The significance of the definition of $\mathbf{n}$ and $\mathbf{m}$
is that it leads to useful bosonic and fermionic operator identities.
For example, one finds that:

\begin{equation}
\,\left\langle \,\widehat{a}_{i}^{\dagger}\widehat{a}_{j}\right\rangle =\left\langle \beta_{i}\alpha_{j}+n_{ij}\right\rangle _{P}\,\,,\end{equation}
 where the weighted average is defined as:

\begin{equation}
\left\langle \widehat{O}\right\rangle =\left\langle O(\overrightarrow{\lambda})\right\rangle _{P}=\,\int O(\overrightarrow{\lambda})\Omega P(\overrightarrow{\lambda},\tau)d\overrightarrow{\lambda}\,.\end{equation}

For representations with \emph{fixed} $n_{ij}$, one thus obtains
a generalized operator-ordering. Classical phase-space distributions
are recovered on setting $\alpha_{i}=\beta_{i}^{*}$, and $n_{ij}=c\delta_{ij}$.
For example, the Glauber-Sudarshan P-representation has $c=0$, while
the Wigner distribution has $c=-1/2$. More generally, this type of
phase space allows for a stochastic covariance, which can dynamically
change in time and space to suit the physical system.

Other useful identities involve the relationship between the action
of operators on the kernel, and the corresponding differential operators
acting on the distribution itself. For simplicity, these are given
in the number-conserving case ($\bm\alpha=\bm\beta=\bm0$, $\bm m=\bm0$,
$\bm m^{+}=\bm0$):

\begin{eqnarray}
\,\widehat{\mathbf{n}}\widehat{\Lambda} & \rightarrow &\mathbf{n}P-\left(\mathbf{I}\pm\mathbf{n}\right)\frac{\overleftrightarrow\partial }{\partial\mathbf{n}}\mathbf{n}P\\
\widehat{\Lambda}\widehat{\mathbf{n}} & \rightarrow &\mathbf{n}P-\mathbf{n}\frac{\overleftrightarrow\partial }{\partial\mathbf{n}}\left(\mathbf{I}\pm\mathbf{n}\right)P \nonumber \,.\end{eqnarray}
where $\left(\overleftrightarrow\partial/\partial{\mathbf n}\right)_{ij} \equiv \overleftrightarrow\partial/\partial n_{ji}$ is a differential operator that acts both to the left and the right.

\subsection{Evolution equations}

There are three main types of problems studied with this approach,
which provides a unified method for interacting fermions and bosons:

\begin{itemize}
\item Canonical ensembles - thermal initial conditions 
\item Quantum dynamics - unitary nonlinear time-evolution 
\item Master equations - open system time-evolution to a steady-state. 
\end{itemize}
The purpose of the phase-space representation is to transform exponentially
complex operator equations into tractable phase-space equations, which
can then be effectively sampled via probabilistic means. For example,
suppose that we wish to calculate a thermal ensemble. The grand-canonical
density operator can be written as an operator differential equation,

\begin{equation}
\frac{d\widehat{\rho}}{d\tau}=-\frac{1}{2}\left[\widehat{H}-\mu\widehat{N}\,,\widehat{\rho}\right]_{+}=\widehat{L}\left[\widehat{\rho}\right]\,\,.\end{equation}

Similarly, one can also treat unitary evolution or evolution under
a master equation as a generalized Liouville operator. By making use
of the operator identities above, and provided conditions of compactness
that allow partial integration are satisfied, one can transform the
exponentially large operator equation into a stochastic equations
that can be treated either numerically or, in some cases, even analytically.
The generic form that results, in the Ito calculus, is:

\begin{eqnarray}
\, d\Omega/\partial t & = & \Omega\left[U+\,\mathbf{g}\,\cdot\bm\zeta\right]\nonumber \\
\, d\bm\lambda/\partial t & = & \mathbf{A}+\mathbf{B}(\bm\zeta-\,\mathbf{g})\,,\end{eqnarray}
 where $\bm\zeta$ is a vector of Gaussian white noises. The function
$\mathbf{g}$ is a `stochastic gauge' function, that can be adjusted
to guarantee the stability of the resulting drift equations.

\begin{figure}
\includegraphics[width=6cm]{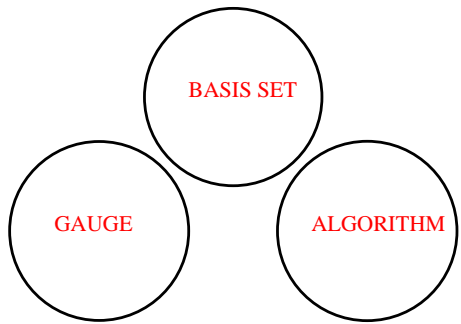}

\caption{Strategies that need to be considered and optimized in quantum simulations.\label{fig:Strategies-that-need}}
\end{figure}

In summary, this method greatly extends the approaches of Glauber,
Sudarshan, Husimi and Wigner. No approximations are needed, apart
from the sampling error, which can be estimated and reduced by using
more samples. The representations use positive, nonsingular distributions
on a relatively small (non-exponential) phase space. This reduces
the overall complexity enormously. The price that is paid is that
many trajectories can be needed to control sampling error, which typically
grows with time. One must also design an appropriate stabilizing gauge
$\mathbf{g}$, as stable trajectories are essential to remove boundary
terms. The overall procedure is outlined schematically in Fig~\ref{fig:Strategies-that-need}.

\section{BOSONS }

The simplest general model of an interacting Bose gas is the Bose-Hubbard
model, which includes nonlinear interactions at each site, together
with linear interactions coupling different sites:

\begin{equation}
\,\widehat{H}(\mathbf{a},\mathbf{a}^{\dagger})=\hbar\left[\sum\sum\omega_{ij}a_{i}^{\dagger}a_{j}+\sum:\widehat{n}_{j}^{2}:\right]\,,\label{eq:Hubbard}\end{equation}
 where the frequency term$\,\omega_{ij}$ is a nonlocal coupling,
which includes chemical potential. The boson number operator is $\,\widehat{n}_{i}=a_{i}^{\dagger}a_{i}$
. The most commonly used technique here is the positive-P representation,
although more general Gaussian methods are also possible.

\subsection{Single-mode phase-diffusion}

As an example, consider the case of a single potential well containing
a BEC in an initial coherent state. After applying the relevant operator
mappings, one obtains the following time-evolution equations:

\begin{eqnarray}
i\,\frac{d\alpha}{d\tau} & = & \left[\text{Re}\left[\beta\alpha\right]+\omega+\sqrt{i}\,\zeta_{1}(\tau)\right]\alpha\nonumber \\
-i\,\frac{d\beta}{d\tau} & = & \left[\text{Re}\left[\beta\alpha\right]+\omega+\sqrt{-i}\,\zeta_{2}(\tau)\right]\beta\nonumber \\
\frac{d\Omega}{d\tau} & = & \Omega\left[\, g_{1}\zeta_{1}(\tau)+g_{2}\zeta_{2}(\tau)\,\right]\,.\end{eqnarray}

Here, unitary evolution leads to nonlinear phase-diffusion, as has
been experimentally observed\cite{Greiner:2002}. The stochastic technique
can be utilized to carry out a simulation of quantum evolution of
an initial coherent state of up to $10^{23}$ bosons!

\begin{figure}
\includegraphics[width=7cm]{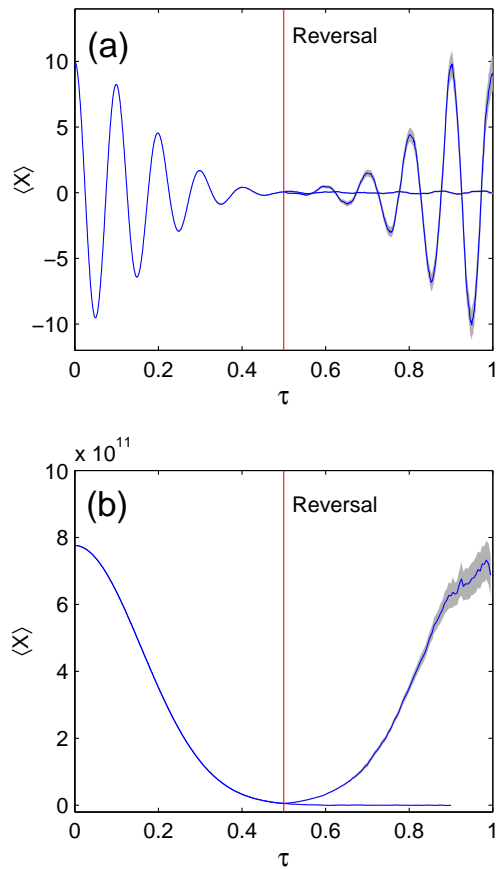} ~

\caption{\label{fig:Simulation-of-100}Simulation of (a) $100$ and (b) $10^{23}$
atoms in a single-mode trap, showing phase-decay together with a recurrence
due to time-reversal.}
\end{figure}

This is shown in Fig.~\ref{fig:Simulation-of-100}, where first $100$
atoms, and then $10^{23}$ atoms were simulated after appropriate
choices of gauges $g_{1,2}$ and noises $\zeta_{1,2}$ \cite{Dowling2005}.
A time-reversal test of unitary evolution was carried out by reversing
the sign of the Hamiltonian, in order to observe a recurrence to the
initial physical state. This is even possible experimentally, using
Feshbach resonances to control the interaction.

\begin{figure}
~~~\includegraphics[width=6cm]{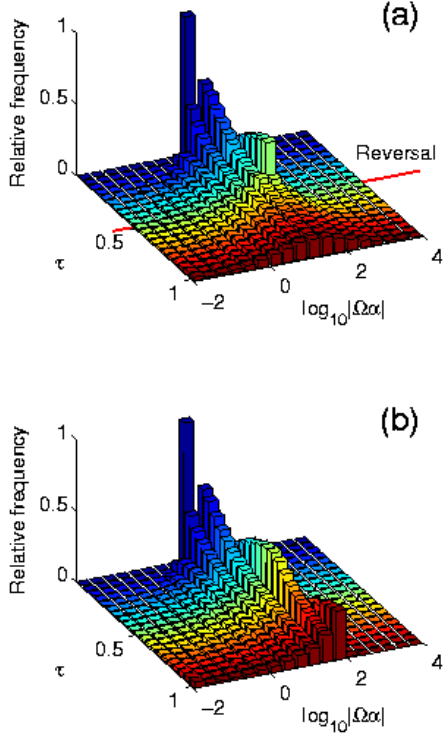}

\caption{\label{fig:Distribution} Phase-space distribution in the single-mode
trap simulations with $N=10$0 showing (a) time-reversal, (b) no time-reversal.}
\end{figure}

The distribution graphs in Fig.~\ref{fig:Distribution} demonstrate
that the mechanism for the recurrence is not through a recurrence
of the entire distribution, as only the physically observable moments
have to show recurrence. The non-uniqueness of the basis means that
the final distribution is actually \emph{different} to the initial
one; the effect of time-reversal is to change the detailed structure
of the diffusive broadening, so that the final and initial distributions
have an equivalent physical density matrix.

\subsection{Optical fibre squeezing experiment}

To a very good approximation, photons in an optical fibre, with the
Kerr nonlinearity present, are an experimental implementation of the
famous one-dimensional Bose gas model in quantum field theory\cite{Drummond1987a,Yurke1989a}.
Phase-space methods were used to make first-principles, testable predictions
of quantum squeezing in this environment. We will show that these
results are in excellent quantitative agreement with experiment, even
including dissipation.

\begin{figure}
\includegraphics[width=7cm]{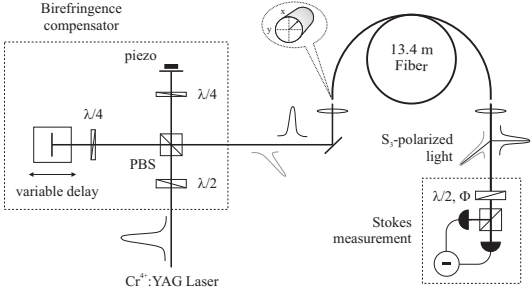}

\caption{\label{cap:Experimental-set-up}Polarisation-squeezing experiment\cite{Heersink2005a}.}
\end{figure}

We focus on recent polarisation squeezing experiments\cite{Heersink2005a},
which are an efficient and flexible method for generating quantum
states in the fibre\cite{Corney2006c}. The experimental set-up is
illustrated in Figure \ref{cap:Experimental-set-up}. Pulses are generated
in pairs and propagate down orthogonal polarisation modes of an optical
fibre. They are then combined in a Stokes measurement of polarisation
squeezing by means of a polarisation rotator, a beam splitter and
two detectors.

Because the experiment involves ultrashort pulses, the quantum description
must use photon-density operators $\widehat{\Psi}_{x}(t,z)$ and $\widehat{\Psi}_{y}(t,z)$
that include a range of spectral components:\begin{eqnarray}
\widehat{\Psi}_{\sigma}(t,z) & \equiv & \frac{1}{\sqrt{2\pi}}\int dk\,\widehat{a}_{\sigma}(t,k)e^{i(k-k_{0})z+i\omega_{0}t}\,,\end{eqnarray}
 where $\sigma=x,y$. The commutation relations of these operators
are $\left[\widehat{\Psi}_{\sigma}(t,z),\widehat{\Psi}_{\sigma'}^{\dagger}(t,z')\right]=\delta(z-z')\delta_{\sigma\sigma'}$.

For convenience, we use scaled variables in propagative frame: $\tau\equiv(t-z/v)$,
$\zeta\equiv z/z_{0}$ and $\widehat{\phi}_{\sigma}\equiv\widehat{\Psi}_{\sigma}\sqrt{vt_{0}/\overline{n}}$,
where $t_{0}$ is the pulse duration, $z_{0}\equiv t_{0}^{2}/|k''|$
is the dispersion length and $2\overline{n}\equiv2|k''|Ac/(n_{2}\hbar\omega_{c}^{2}t_{0})$
is the photon number in a soliton pulse.

To describe the evolution of the photon flux $\widehat{\phi}_{\sigma}(\tau,\zeta)$,
we employ a quantum model of a radiation field propagating along a
silica fibre, including $\chi^{(3)}$ nonlinear responses of the material
and non-resonant coupling to phonons\cite{Drummond2001a,Carter1991a}.The
phonons provide a non-Markovian reservoir that generates additional,
delayed nonlinearity, as well as spontaneous and thermal noise. Because
of fibre birefringence, the two polarisation components do not temporally
overlap for most of the fibre length, and so the cross-polarisation
component of the Raman gain is neglected. The result, after discretization,
is a Hubbard model like Eq.(\ref{eq:Hubbard}), except with additional
coupling to phonon reservoirs.

The quantum operator equations are obtained by integration of the
Heisenberg equations for the phonon operators to derive quantum Langevin
equations for the photon-flux field:\begin{eqnarray}
\frac{\partial}{\partial\zeta}\widehat{\phi}_{\sigma}(\tau,\zeta) & = & \frac{i}{2}\frac{\partial^{2}}{\partial\tau^{2}}\widehat{\phi}_{\sigma}(\tau,\zeta)+i\widehat{\Gamma}_{\sigma}(\tau,\zeta)\widehat{\phi}_{\sigma}(\tau,\zeta)\label{eq:Raman-operator}\\
 & + & i\int_{-\infty}^{\infty}d\tau'h(\tau-\tau')\widehat{\phi}_{\sigma}^{\dagger}(\tau',\zeta)\widehat{\phi}_{\sigma}(\tau',\zeta)\widehat{\phi}_{\sigma}(\tau,\zeta).\nonumber \end{eqnarray}
 where the nonlinear response function $h(\tau)$ includes contributions
from both the instantaneous electronic response and the Raman response
determined by the gain function $\alpha^{R}(\omega)$\cite{Stolen1984a,Stolen1989a,Drummond2001a}.
The correlations of the reservoir fields are:\begin{eqnarray}
\left\langle \widehat{\Gamma}_{\sigma}^{\dagger}(\omega',\zeta')\widehat{\Gamma}_{\sigma'}(\omega,\zeta)\right\rangle  & = & \frac{\alpha^{R}(|\omega|)}{\overline{n}}\left[n_{\mathrm{th}}(|\omega|)+\Theta(-\omega)\right]\nonumber \\
 &  & \times\delta(\zeta-\zeta')\delta(\omega-\omega')\delta_{\sigma\sigma'}\,\,,\end{eqnarray}
 where $n_{{\textrm{th}}}$ is the temperature-dependent Bose distribution
of phonon occupations. The Stokes ($\omega<0$) and anti-Stokes ($\omega>0$)
contributions to the Raman noise are included by means of the Heaviside
step function $\Theta$.

In all, we have over $10^{8}$ photons in more than $10^{2}$ modes,
corresponding to an enormously large Hilbert space. Quantum dynamical
simulations of such systems have been performed exactly using the
$+P$ representation\cite{Carter1987a,Drummond1987a}. However, for
large photon number $\overline{n}$ and short propagation distance
$L$, these exact squeezing predictions agree with a truncated Wigner
phase-space method\cite{Drummond1993a}, which allows faster calculations.
In effect, the Wigner representation maps a field operator to a stochastic
field: $\widehat{\phi}_{\sigma}(\zeta,\tau)\rightarrow\phi_{\sigma}(\zeta,\tau).$
Stochastic averages involving this field then correspond to symmetrically
ordered correlations of the quantum system. Because of the symmetric-ordering
correspondence, quantum effects enter via vacuum noise. 

After the mapping, we obtain a Raman-modified stochastic nonlinear
Schroedinger equation for the photon flux that is of exactly the same
form as Eq.~(\ref{eq:Raman-operator})\cite{Drummond2001a,Carter1995a}.
The correlations of the Raman noise fields $\Gamma_{\sigma}$ and
the initial vacuum noise are, respectively, \begin{eqnarray}
\left\langle \Gamma_{\sigma}(\omega,\zeta)\Gamma_{\sigma'}(\omega',\zeta')\right\rangle  & = & \frac{\alpha^{R}(|\omega|)}{\overline{n}}\left[n_{\mathrm{th}}(|\omega|)+\frac{1}{2}\right]\nonumber \\
 &  & \times\delta(\zeta-\zeta')\delta(\omega-\omega')\delta_{\sigma\sigma'}\,,\nonumber \\
\left\langle \Delta\phi_{\sigma}(\tau,0)\,\Delta\phi_{\sigma'}^{*}(\tau',0)\right\rangle  & = & \frac{1}{2\overline{n}}\delta(\tau-\tau')\delta_{\sigma\sigma'}\,.\end{eqnarray}
 Because of the symmetrically ordered mapping, the Stokes and anti-Stokes
contributions to the Wigner Raman noise are identical.

\begin{figure}
\includegraphics[width=7cm]{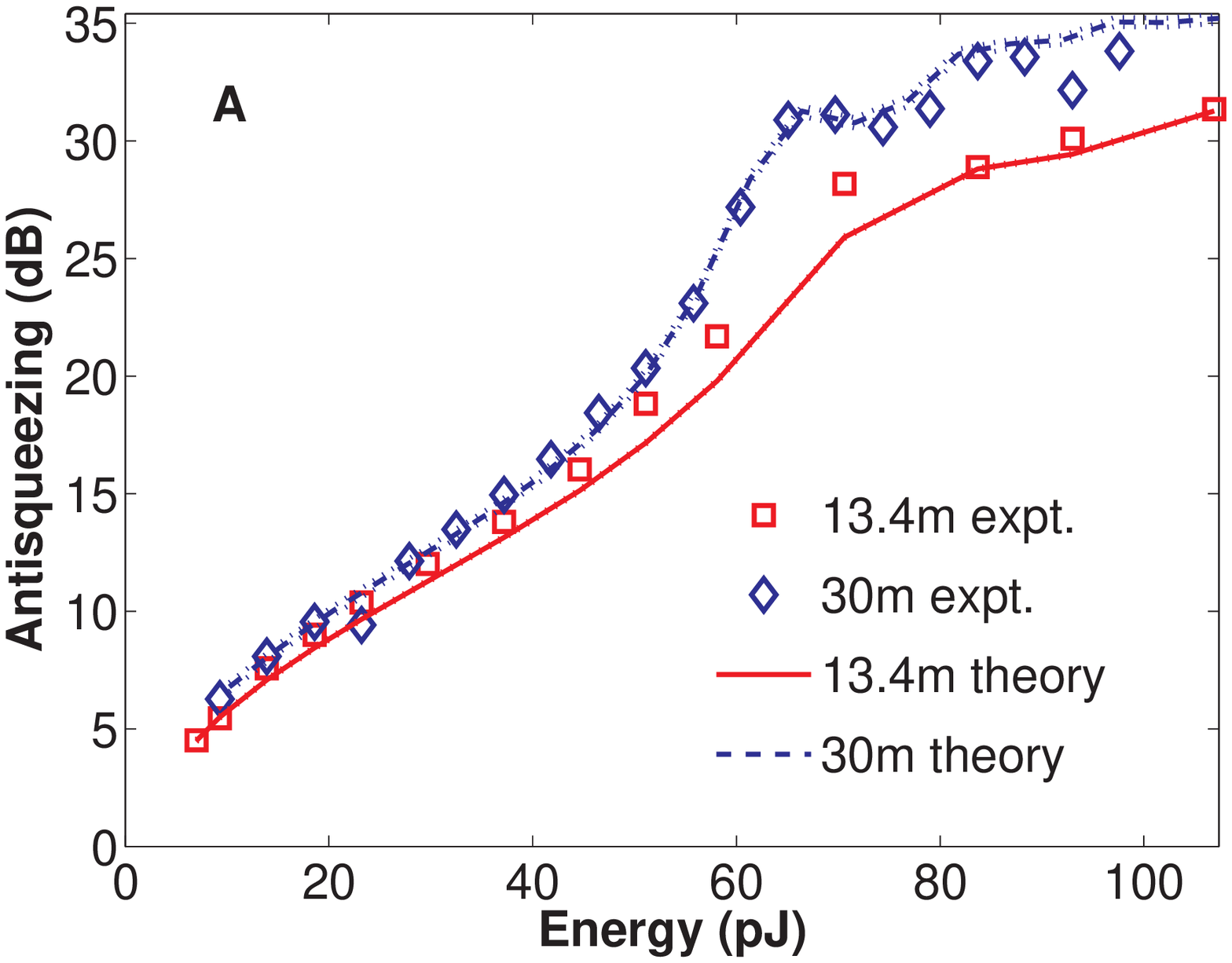} \includegraphics[width=7cm]{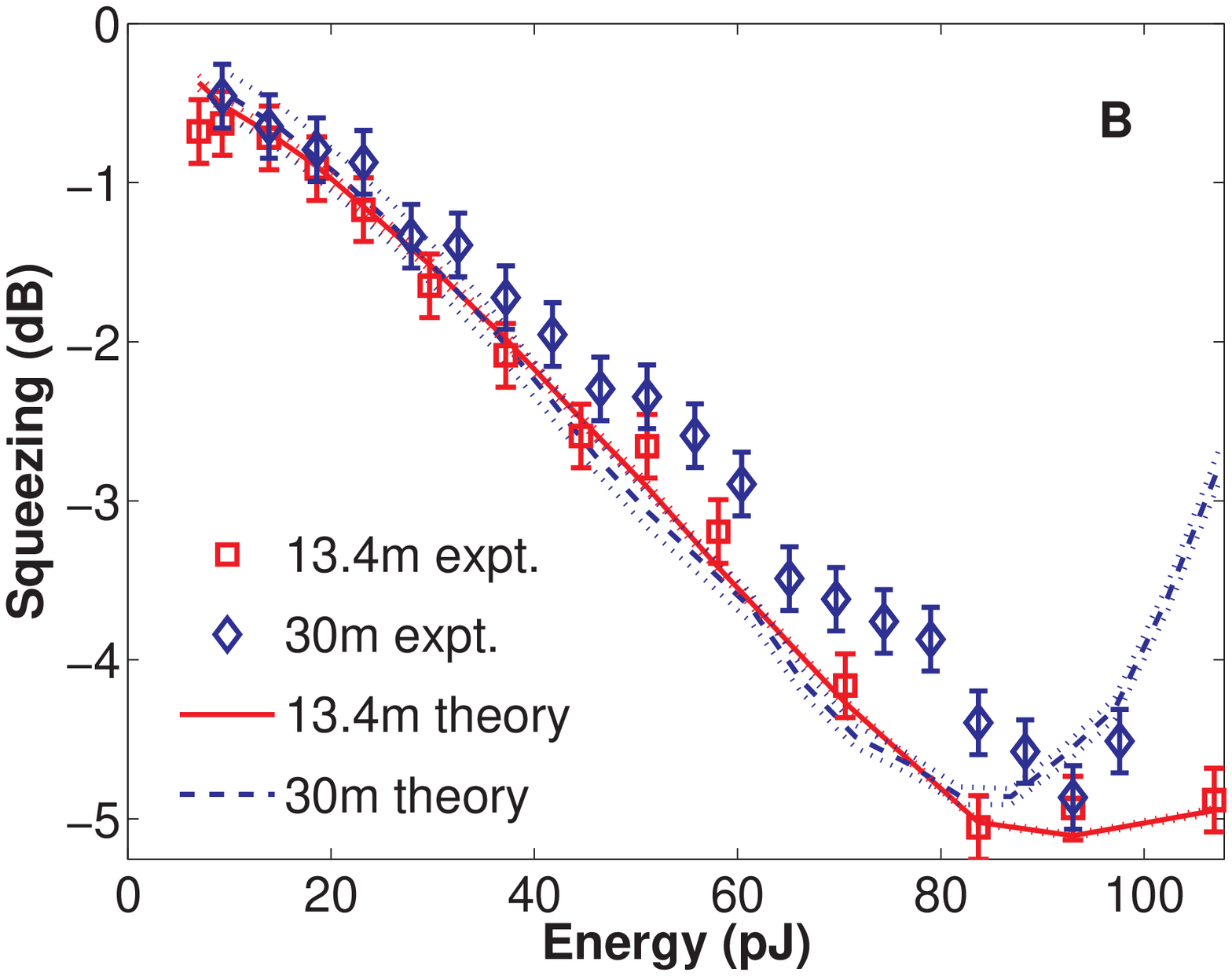}

\caption{\label{cap:antisqueezing-and-squeezing}Antisqueezing and squeezing
for $L_{1}=13.4{\textrm{m}}$ (squares) and $L_{2}=30{\textrm{m}}$
(diamonds) fibres. Solid and dashed lines show the simulation results
for $L_{1}=13.4{\textrm{m}}$ and $L_{2}=30{\textrm{m}}$, respectively.
Dotted lines indicate sampling error in simulation results. Simulations
are adjusted for linear loss of $24\%$ and low-frequency GAWBS noise,
which mainly affects the squeezing only at low power. Parameters are
parameters: $t_{0}=74{\textrm{fs}}$, $z_{0}=0.52{\textrm{m}}$, $\overline{n}=2\times10^{8}$,
$E_{s}=54{\textrm{pJ}}$ and $\lambda_{0}=1.51\mu{\textrm{m}}$.}
\end{figure}

Antisqueezing and squeezing results are shown figure \ref{cap:antisqueezing-and-squeezing},
for 13.4m and 30m of fibre, with and without the excess phase noise
included. The theoretical results for both squeezing and antisqueezing
closely match the experimental data. The results also show a deterioration
of squeezing at higher intensity due to Raman effects, especially
for longer fibre lengths.


\subsection{BEC collision with 150,000 atoms from first principles}

\newcommand{\op}[1]{\widehat{#1}} \newcommand{\dagop}[1]{\widehat{#1}^{\dagger}}
The collision of pure ${}^{23}$Na BECs, as in a recent experiment
at MIT\cite{Vogels2002}, represents another opportunity for observational
tests of first-principles quantum dynamical simulations\cite{Deuar2005,Deuar2006b}.
In the simulations, a $1.5\times10^{6}$ atom condensate is prepared
in a cigar-shaped magnetic trap with frequencies 20 Hz axially in
the {}``X'' direction, and 80 Hz radially ({}``Y'' and {}``Z'').
A brief Bragg laser pulse coherently imparts an X velocity of $2{\rm v_{Q}}\approx20$
mm/s to half of the atoms, which is much greater than the sound velocity
of $3.1$ mm/s. Another much weaker pulse generates a small 2\% \char`\"{}seed\char`\"{}
wavepacket at a Y velocity of ${\rm v}_{s}=9.37$ mm/s relative to
the center of mass.

At this point the trap is turned off so that the wavepackets collide
freely. In a center-of-mass frame, atoms are scattered preferentially
into a spherical shell in momentum space with mean velocities ${\rm v_{s}}\approx{\rm v_{Q}}$.
Simultaneously, a four-wave mixing process generates a new coherent
wavepacket at Y velocity -${\rm v_{s}}$, as well as growing the strength
of both of the wavepackets at $\pm{\rm v_{s}}$ by Bose enhanced scattering.

\begin{figure}
\includegraphics[width=8cm]{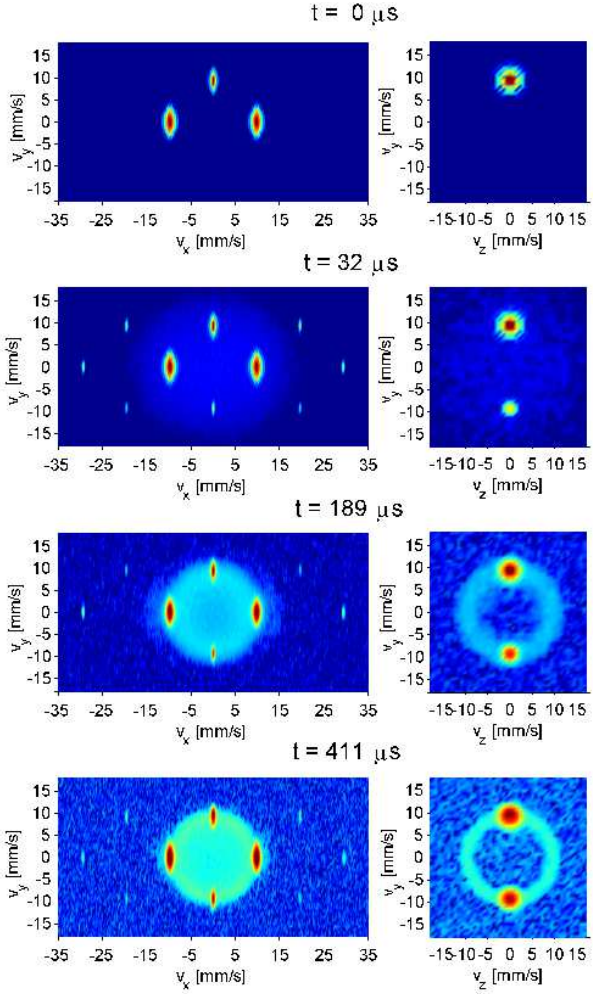}~~

\caption{\label{nkxky} \textit{Momentum space snap-shots} in the center-of
mass frame. \textbf{left:} Velocity distributions in the axial (x)
and radial (y) directions. \textbf{right:} Radial distribution at
x=0. The formation of the fourth coherent wavepacket at ${\rm v_{y}\approx-v_{s}}$
and the scattered shell at $|v|\approx{\rm v_{s}=9.37mm/s}$ are seen.
Logarithmic color scale. Average of 1492 trajectories.}
\end{figure}

\begin{figure}
\includegraphics[width=8.5cm]{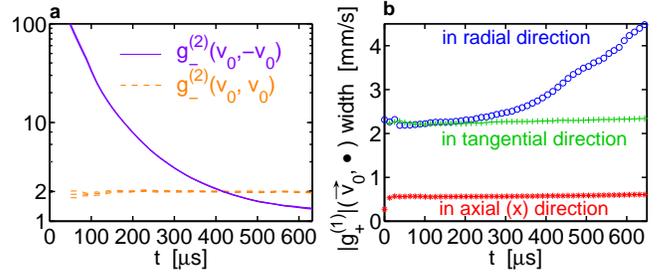}

\caption{\label{g2t}\textit{Correlations between scattered atoms: time evolution}.
Plate \textbf{a} shows the extremely strong number correlations $g^{(2)}({v_{0},-v_{0}})$
between atoms with opposite velocity (solid line) in the scattered
shell at $|{v_{0}}|={\rm v_{s}}$ (away from the coherent wavepackets),
and thermal correlations $g^{(2)}({v_{0},v_{0}})=2$ between scattered
atoms at the same velocity (dashed). Triple lines indicate uncertainty.
Plate \textbf{b} shows the coherence width in velocity space for scattered
atoms at similar velocities centered around ${v_{0}}$. Plotted is
the Full-width at half-maximum (FWHM) of $|g^{(1)}({v_{0},v_{0}+v})|$. }
\end{figure}

The dynamics of the distribution of atom velocities and correlations
between the scattered atoms have been calculated and are shown in
Figures~\ref{nkxky} and~\ref{g2t}. Such correlations have recently
become experimentally measurable\cite{Greiner2005,Folling2006,Schellekens2006},
and correlation behaviour qualitatively similar to that predicted
by this model have been seen\cite{Westbrooke2006}. The system is
described by: \begin{equation}
\op{H}=\int\,\left[\frac{\hbar^{2}}{2m}\nabla\dagop{\Psi}\nabla\op{\Psi}+\frac{g}{2}\op{\Psi}^{\dagger2}\op{\Psi}^{2}\right]\, d\,^{3}\vec{x}\label{H}\end{equation}
 The simulation is carried out using the positive-P representation
in the center-of-mass frame from the moment the lasers and trap are
turned off \mbox{($t=0$).} The initial wavefunction is modeled
as the coherent-state mean-field Gross-Pitaevskii (GP) solution of
the trapped $t<0$ condensate, but modulated with a factor $\left[\sqrt{0.49}e^{im{\rm v_{Q}}{\rm x}/\hbar}+\sqrt{0.49}e^{-im{\rm v_{Q}}{\rm x}/\hbar}+\sqrt{0.02}e^{-im{\rm v_{s}}{\rm y}/\hbar}\right]$
which imparts the initial velocities. The field Hamiltonian is discretized
with a lattice size of $432\times105\times50$, again generating a
Hubbard-type Hamiltonian like Eq (\ref{eq:Hubbard}).

As one might expect of a method that attacks such an exponentially
complex problem, there are limitations. Most significantly, the size
of the sampling uncertainty grows with time, and eventually reaches
a size where it is no longer practical to produce enough trajectories
to get useful precision. In the above case a useful observable-to-noise
ratio lasted until $t\lesssim410\,\mu$s. In general the simulation
time possible depends on several factors: coarser lattices, weaker
interactions, or smaller density all extend it. This time can be estimated
using the formulae found in \cite{Deuar2006a}. Comparisons were made
with a previous approximate simulation\cite{Norrie2005}, using a
truncated Wigner method\cite{Drummond1993a,Wigner1932a}. The approximate
method was less accurate at large momentum cutoff, due to a diverging
truncation error.

The model treats $M=2.268\times10^{6}$ interacting momentum modes.
Since each of the $N=1.5\times10^{5}$ atoms can be in any one of
the modes, the Hilbert space contains about $N_{s}\approx M^{N}\approx10^{1,000,000}$
orthogonal quantum states. In terms of accessible states at fixed
number, there are $N_{s}\approx(M/N)^{N}\approx2^{600,000}$ states,
or $600,000$ qubits.

\textbf{This is the largest Hilbert space ever treated in a first-principles
quantum dynamical simulation.}

\section{FERMIONS}

\subsection{Hubbard model}

To demonstrate the utility of this fermionic representation, we next
consider the \emph{fermionic} Hubbard model\cite{Hubbard1963a}. This
is well-known in condensed matter physics as the simplest model of
interacting fermions on a lattice:

\begin{equation}
\widehat{H}(\widehat{\mathbf{n}}_{\downarrow},\widehat{\mathbf{n}}_{\uparrow})=-\sum_{ij,\sigma}t_{ij}\widehat{n}_{ij,\sigma}+U\sum_{j}:\widehat{n}_{j,j,\downarrow}\widehat{n}_{j,j,\uparrow}:\,\,.\end{equation}
 Here $\widehat{n}_{ij,\sigma}\equiv\widehat{a}_{i,\sigma}^{\dagger}\widehat{a}_{j,\sigma}$,
for lattice index $j$ and spin index $\sigma=(\uparrow,\downarrow)=(-1,1)$,
while $t_{ij}$ is the inter-site coupling and $U$ is the strength
of on-site interaction between particles.

Thought to be relevant to high-$T_{c}$ superconductors\cite{Schrieffer1964a},
the Hubbard model has had renewed interest because it describes an
ultra-cold gas in an optical lattice\cite{Hofstetter2002a}, as has
been experimentally realised by Köhl et al\cite{Kohl2005a}. Within
this simple model is a great complexity, that leads to sampling error
problems for quantum Monte Carlo (QMC) methods because of negative
weights\cite{Linden1992a}. Such sign problems occur for repulsive
interactions away from half-filling in two or more dimensions, and
increase with lattice size and interaction strength\cite{Santos2003a}.

\begin{figure}
\includegraphics[width=7cm]{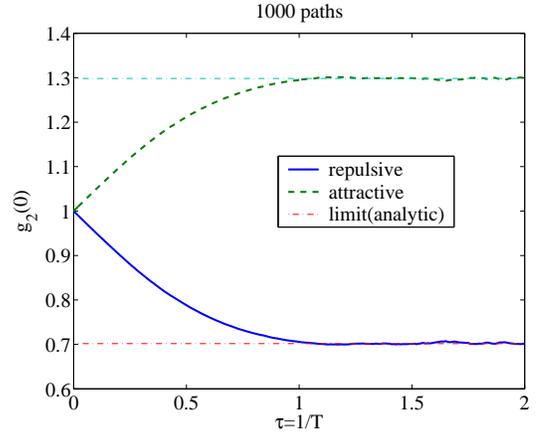}

\caption{\label{cap:1D-Hubbard-model:}1D Hubbard model: Correlation function
$\left\langle n_{\downarrow}n_{\uparrow}\right\rangle $ versus temperature.The
100-site numerical solution is compared with the zero-temperature
exact solution of an infinite lattice\cite{Lieb1968a}: $t=1$ and
$U=2$.}
\end{figure}

\begin{figure}
\includegraphics[width=7cm]{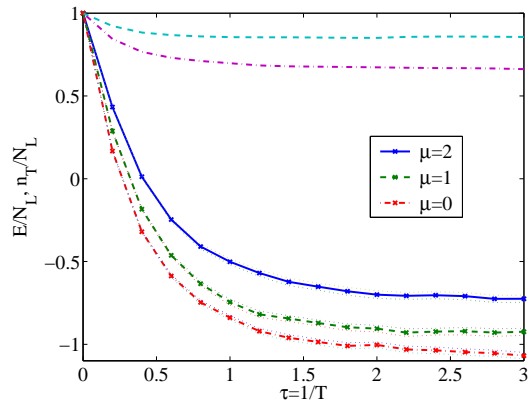}

\caption{\label{cap:2D-Hubbard-model}2D Hubbard model on a 16$\times$16
lattice: Energy as a function of of temperature for various chemical
potentials. $t=1$ and $U=4$.}
\end{figure}

Results of recent phase-space numerical simulations\cite{Corney2004a}
in one and two dimensions are shown in Figures $\ref{cap:1D-Hubbard-model:}$
and \ref{cap:2D-Hubbard-model}. The sampling error remains well-controlled
at low temperatures, even for filling factors in 2D for which other
QMC methods suffer sign problems.

To explain the method used, we first note that the Hubbard Hamiltonian
conserves number, so the number-conserving subset of Gaussian operators
provides a complete basis, i.e. the anomalous variables remain zero.
The mappings given above can be applied to the grand canonical equilibrium
equation to give the It\^{o} phase-space equations\cite{Corney2006b}:

\begin{eqnarray}
\frac{d\mathbf{n}_{\sigma}}{d\tau} & = & \frac{1}{2}\left\{ \left(\mathbf{I}-\mathbf{n}_{\sigma}\right)\bm T_{\sigma}^{(1)}\!\mathbf{n}_{\sigma}+\mathbf{n}_{\sigma}\bm T_{\sigma}^{(2)}\!\left(\mathbf{I}-\mathbf{n}_{\sigma}\right)\right\} .\ \end{eqnarray}
 The propagation matrices are defined for $U>0$, as\begin{eqnarray}
T_{i,j,\sigma}^{(r)} & = & t_{ij}-\delta_{i,j}\left\{ Un_{j,j,-\sigma}-\mu+\sigma\xi_{j}^{(r)}\right\} ,\end{eqnarray}
 where the stochastic terms are Gaussian white noises with the correlations

\begin{eqnarray}
\left\langle \xi_{j}^{(r)}(\tau)\,\xi_{j'}^{(r')}(\tau')\right\rangle  & = & 2U\delta(\tau-\tau')\delta_{j,j'}\delta_{r,r'}\,\,.\end{eqnarray}
 Associated with each stochastic path is a weight, governed by $d\Omega/d\tau=-\Omega H(\mathbf{n}_{1},\mathbf{n}_{-1}).$
Importantly, because the choice of mapping, the phase-space equations
are real and the weights thus remain positive, avoiding the usual
manifestation of the sign problem.

More precise numerical simulations by Assaad et al\cite{Assaad2005a}
have revealed that there is difficulty in sampling ground state properties
with these phase-space equations. However, they also show that the
correct ground-state results can be obtained by a supplementing the
phase-space simulations with a symmetry projection procedure.

Finally, we remark that the mapping from the Hubbard model to phase-space
equation is far from unique. Thus these phase-space simulations of
the Hubbard model may well be improved by appropriate choice of basis
subset and stochastic gauge, as for bosonic simulations.

\section{Summary}

In summary, coherence theory and coherent-state methods leads to a
unified phase-space representation for bosonic and fermionic quantum
many-body systems, which are useful in simulations both in real time
and in inverse temperature. Calculations have been carried out in
one, two and three dimensions, with up to $10^{23}$ particles and
$10^{6}$ modes. This is equivalent to a Hilbert space of nearly a
million qubits. Phase-space ideas are also applicable to other complex
systems\cite{Gardiner:1977,Drummond:2004} - ranging from genetics,
astrophysics, and biochemistry, to condensed matter, particle physics
and possibly even molecular physics.

\emph{Acknowledgements:} We acknowledge funding from the Australian
Research Council and useful discussions with K.V. Kheruntsyan.  

\bibliography{Phasespace}
 \bibliographystyle{apsrev}
\end{document}